\pdfoutput=1
\RequirePackage{amsmath}
\documentclass[a4paper,fleqn,usenatbib]{mnras}

\usepackage[T1]{fontenc}
\usepackage{ae,aecompl}

\usepackage{graphicx}	
\usepackage{amssymb}	

\title[Dating boxy/peanut bulge formation]{Observational constraints to boxy/peanut bulge formation time}

\author[I. P\'erez]{
I. P\'erez,$^{1,2}$\thanks{E-mail: isa@ugr.es}
I. Mart\'inez-Valpuesta,$^{3,4}$
T. Ruiz-Lara,$^{1,2,3,4}$
A. de Lorenzo-Caceres,$^{5}$
\newauthor J. Falc\'on-Barroso,$^{3,4}$
E. Florido,$^{1,2}$
R.M. Gonz\'alez Delgado,$^{6}$
M. Lyubenova,$^{7}$
\newauthor R.A. Marino,$^{8}$
S.F. S\'anchez,$^{5}$
P. S\'anchez-Bl\'azquez,$^{9,10}$
G. van de Ven,$^{11}$
A. Zurita$^{1,2}$
\\
$^{1}$Departamento de F\'isica Te\'orica y del Cosmos, Universidad de Granada, Campus de Fuentenueva, E-18071 Granada, Spain \\
$^{2}$Instituto Carlos I de F\'isica Te\'orica y Computacional, Universidad de Granada, E-18071 Granada, Spain\\
$^{3}$Instituto de Astrof\'isica de Canarias, Calle V\'ia L\'actea s/n, E-38205 La Laguna, Tenerife, Spain\\
$^{4}$Universidad de La Laguna, Dpto. Astrof\'isica, E-38206 La Laguna, Tenerife, Spain\\
$^{5}$Instituto de Astronom\'ia, Universidad Nacional Aut\'onoma de M\'exico, A.P. 70-264, 04510, M\'exico, D.F.\\
$^{6}$Instituto de Astrof\'isica de Andaluc\'ia (CSIC), PO Box 3004, 18080, Granada, Spain \\
$^{7}$Kapteyn Astronomical Institute, University of Groningen, Landleven 12, NL-9747 AD Groningen, the Netherlands\\
$^{8}$Institute for Astronomy, ETH Zurich, Wolfgang-Pauli-Strasse 27, 8093 Zurich, Switzerland\\
$^{9}$Departamento de F\'isica Te\'orica, Universidad Aut\'onoma de Madrid, Cantoblanco E-28049, Spain\\
$^{10}$Instituto de Astrof\'isica, Universidad Pont\'ifica Cat\'olica de Chile, Av. Vicu\~na Mackenna 4860, Santiago, Chile\\
$^{11}$Max Planck Institute for Astronomy, K\"onigstuhl 17, D-69117 Heidelberg, Germany\\
}
 
\date{Accepted XXX. Received YYY; in original form ZZZ}

\pubyear{2015}

\begin{document}
\label{firstpage}
\pagerange{\pageref{firstpage}--\pageref{lastpage}}
\maketitle

\begin{abstract}
Boxy/peanut bulges are considered to be part of the same stellar structure as bars and both could be linked through the buckling instability. The Milky Way is our closest example. The goal of this letter is determining if the mass assembly of the different components leaves an imprint in their stellar populations allowing to estimate the time of bar formation and its evolution. To this aim we use integral field spectroscopy to derive the stellar age distributions, SADs, along the bar and disc of NGC~6032. The analysis shows clearly different SADs for the different bar areas. There is an underlying old (~$\geqslant~12$~Gyr) stellar population for the whole galaxy. The bulge shows star formation happening at all times. The inner bar structure shows stars of ages older than 6 Gyrs with a deficit of younger populations. The outer bar region presents a SAD similar to that of the disc. To interpret our results, we use a generic numerical simulation of a barred galaxy. Thus, we constrain, for the first time, the epoch of bar formation, the buckling instability period and the posterior growth from disc material. We establish that the bar of NGC~6032 is old, formed around 10 Gyr ago while the buckling phase possibly happened around 8 Gyr ago. All these results point towards bars being long-lasting even in the presence of gas.

\end{abstract}




\section{Introduction}

In the last years, observation and theory have converged towards secular evolution, linked to the bar formation, as the main mechanism for the formation of the Milky Way boxy bulge \citep[e.g.,][]{2010ApJ...720L..72S,2011ApJ...734L..20M,2016arXiv161109023D}. This explanation invokes a buckling instability of the bar, some time after the bar forms, to create the central `bulgy' structure. Bulges formed in this way have been previously related to the boxy/peanut (B/P) bulges observed in external galaxies \citep[e.g.,][]{1981A&A....96..164C, 2006ApJ...645..209D}. Roughly, 40\% of disc galaxies present these type of bulges \citep{2000A&AS..145..405L} at z=0.  A recent work \citep{2016ApJ...825L..30E} argues that most of the B/P bulges have been formed through a bar buckling instability. 
The moment at which the bar forms would constrain the formation time of the B/P bulge and help us understand the effects of secular evolution on the structures of the Milky Way as well as in other disc galaxies. 
Dating the time of bar formation is challenging as the stars currently populating the bar are not necessarily coeval with the formation of the bar. However, simulations suggest that there should be a period of intense star formation during the formation of the bar, lasting as long as there is material available to form stars \citep{1994ApJ...430L.105F}. This star formation should leave its imprint in the stellar content of the galaxy. 

The analysis of the spatially resolved star formation history offers a unique window to explore the formation of galactic structures. 
Accurate star formation histories (SFHs) and age-metallicity relations are fundamental to understand the build-up of the stellar component and the enrichment of the interstellar medium of galaxies of different types and in different environments. Very few studies have analysed the bar stellar properties in detail due to the high quality of the data required for such purposes. In addition, most of them are mainly based on line-strength indices \citep[e.g.,][]{2006ApJS..163..270G, 2007A&A...465L...9P, 2009A&A...495..775P}. Modern full spectrum fitting techniques \citep[][]{2006MNRAS.365...74O,2005MNRAS.358..363C,2009A&A...501.1269K,2016RMxAA..52...21S} allow us to obtain, not only mean ages and metallicities, but also stellar age distributions shaping the observed spectra.

In this letter we present the analysis of the stellar age distributions (SADs) of different regions of the galaxy NGC~6032 for which we have integral field spectroscopic CALIFA data \citep[][]{2012A&A...538A...8S}. NGC~6032 is a spiral galaxy classified as SBb with a total stellar mass of $2.6 \times 10^{10}M_{\odot}$ as derived from the SDSS colours \citep{2013A&A...554A..58S}. The analysed morphological structures comprise the bulge, the bar and the disc, focusing particularly on the fine-structures of the bar. The bar in this galaxy shows a barlens structure, as described in \citet{2011MNRAS.418.1452L}, and some light enhancements at the ends of it (see Fig.~\ref{fig:light}), and therefore it is a perfect candidate to explore bar evolution. The aim is to determine whether we can observe stellar populations signatures of how these different structures formed, and to compare them with results from a state-of-art numerical simulation of bar formation.

\section{Data}

The CALIFA project \citep[][]{2012A&A...538A...8S} has observed more than 700 galaxies from the local Universe with the PMAS/PPAK instrument \citep{2005PASP..117..620R} mounted at the $3.5$ m telescope at Calar Alto. Three different exposures were taken for each object following a dithered scheme (to reach a filling factor of 100$\%$) in two observational setups (V1200 and V500). The total exposure time for the V500 setup (wavelength range 3745--7300 \AA, R $\sim$ 850) is 2700 s and 5400 s for the V1200 (3400--4750 \AA, R $\sim$ 1650) setup. The diameter of each spaxel is 2.7" which corresponds,for NGC~6032, to a physical size of 805~pc given a comoving distance to the galaxy of 62.5~Mpc. For more information about the quality of the data and the data reduction procedure (v1.4) see \citet{2013A&A...549A..87H} and \citet{2014arXiv1409.8302G}. We also make use of the $g$-band SDSS image, using the SDSS seventh Data Release \citep[DR7][]{2009ApJS..182..543A}, to obtain the radial light distribution of NGC~6032 and to define its different morphological regions (see Fig.~\ref{fig:light}). 

\section{Stellar population analysis}

We analyse the stellar content in NGC~6032 from the CALIFA data using a method based on full-spectrum fitting techniques \citep[][]{2011MNRAS.415..709S, 2014A&A...570A...6S, 2015MNRAS.446.2837S, 2016MNRAS.456L..35R} that has been proven successful at replicating results obtained from the more reliable analysis of Colour-Magnitude Diagrams \citep[][]{2015A&A...583A..60R}. In short:


(i) We correct the observed datacubes for the stellar kinematics effect in order to have every spectrum in the datacubes rest framed and at a common spectral resolution (8.4~\AA). We use the results from a stellar kinematics extraction method designed for dealing with CALIFA data \citep[][]{2017A&A...597A..48F} that applies an adaptive Voronoi binning \citep[][]{2003MNRAS.342..345C} with a S/N goal of 20, using only spaxels with continuum S/N~$>$~3. Finally, it makes use of the penalised pixel fitting code \citep[{\tt pPXF};][]{2004PASP..116..138C, 2011MNRAS.413..813C} to extract the stellar kinematics.

(ii) Once the original CALIFA data are at rest frame and at a common spectral resolution of 8.4~\AA, we integrate over elliptical annuli (on the unbinned, kinematics-corrected datacubes) using different ellipticity and position angle (PA) values and variable widths to reach a minimum S/N of 20 (per~\AA), see Fig.~\ref{fig:light}. These values were derived by fitting successive ellipses to the galaxy isophotal light distribution in the $g$-band image using variable ellipticities and PA and fixing their centre.

(iii) We subtract the emission from the gaseous component (emission lines) to the radially-integrated spectra using {\tt GANDALF} \citep[Gas AND Absorption Line Fitting;][]{2006MNRAS.366.1151S, 2006MNRAS.369..529F}.  We use the \citet[][]{2010MNRAS.404.1639V} (V10 hereafter) models based on the MILES library\footnote{The models are publicly available at \url{http://miles.iac.es}} \citep{2006MNRAS.371..703S} as stellar templates.  

(iv) Finally, we recover the light- and mass-weighted SADs by applying the {\tt STECKMAP} \citep[STEllar Content and Kinematics via Maximum A Posteriori likelihood;][]{2006MNRAS.365...74O, 2006MNRAS.365...46O} code to the emission-cleaned spectra. {\tt STECKMAP} is able to simultaneously recover the stellar content and stellar kinematics using a Bayesian method via a maximum {\it a posteriori} algorithm. We also use the entire set of the V10 models while running {\tt STECKMAP} and fix the stellar kinematics to the values computed with {\tt pPXF} to avoid the metallicity-velocity dispersion degeneracy \citep[][]{2011MNRAS.415..709S}. The smoothing parameters used in this work are $\mu_x$ $=$ 0.01 and $\mu_Z$ $=$ 100 for the SAD and the age-metallicity relation, respectively. We choose to use the full age range present in the V10 models to avoid biasing the outcome. Although this can lead to outputs of ages older than the age of the Universe, this is the usual procedure in this type of studies \citep[e.g.][]{2015MNRAS.446.2837S}.

The radially-resolved SADs are presented in Fig.~\ref{fig:sfh} in three different ways. The left and middle panels show the light- and mass-weighted SADs, respectively. The SADs at each radius are normalised to the total light or mass within each ellipse, i.e. the sum of the light or mass fraction at each radius is 1. This visualisation allows us to properly compare the SAD in the bulge, bar, and disc regions avoiding artefacts due to the different values of the surface brightness or surface mass density of each component. On the other hand, the right hand panel shows the average SAD for each region with the corresponding propagated errors. These error in the SADs are computed by means of 25 Monte Carlo simulations. Once {\tt STECKMAP} has determined the best combination of model templates to fit the observed spectrum, we add noise based on the spectrum S/N and run {\tt STECKMAP} again. This test is done 25 times and the error in the SAD is considered as the standard deviation of the recovered light and mass fractions at each age.

\section{The stellar content of NGC~6032}

\begin{figure*}
	\includegraphics[width=0.75\textwidth]{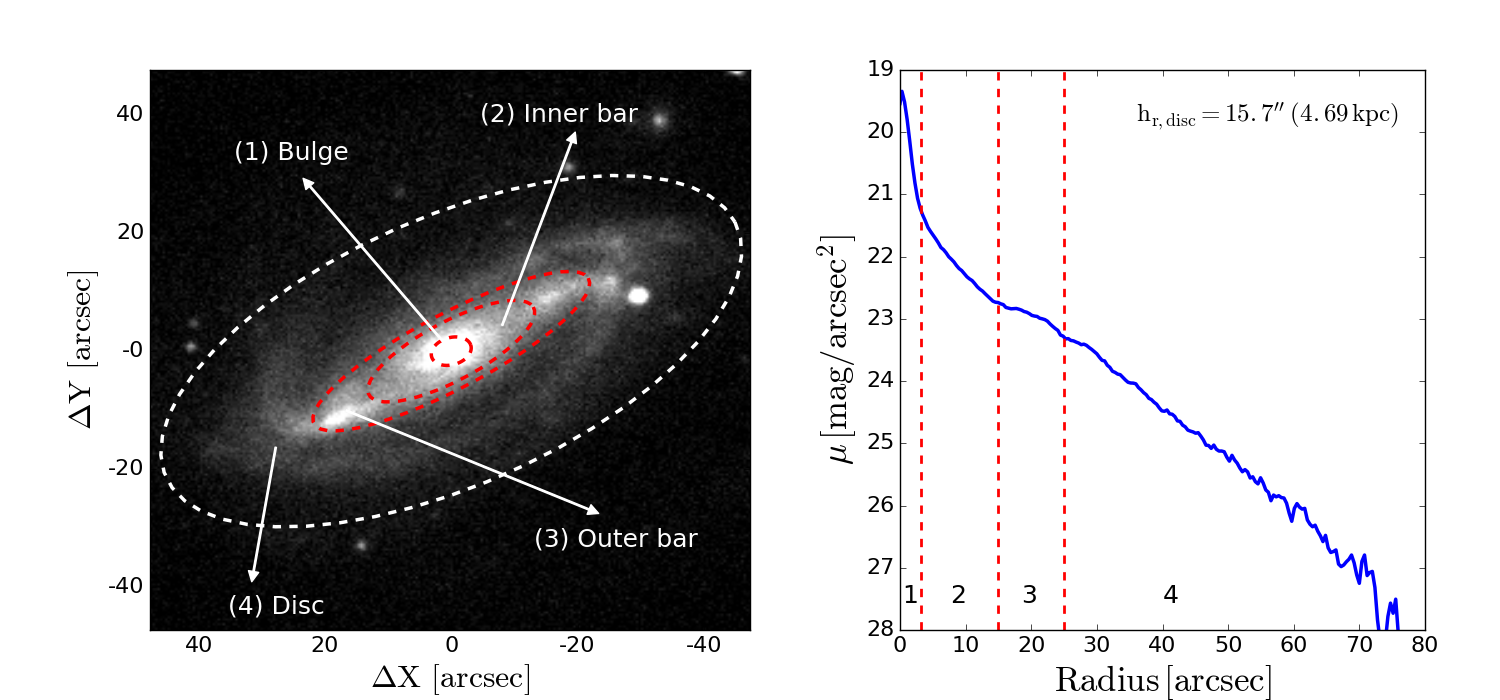}
    \caption{Left panel: $g$-band image of NGC 6032 from SDSS with the analysed regions overplotted. The white dashed line represents the position angle and ellipticity of the disc. Right panel: Radial surface $g$-band brightness distribution with the different regions, dashed red lines, with numbering from the previous panel. The disc scale-length is shown in the upper-right corner. }
    \label{fig:light}
\end{figure*}

Before describing the results obtained in the stellar population analysis we will first define the structural regions distinguished in this work (see Fig.~\ref{fig:light}). 

(i) Region 1: The bulge region. Radially, the inner 3 arcsec. A central light enhancement characterised by an excess of light from the exponential disc in the light distribution.

(ii) Region 2: The inner bar. Radially comprised between 3 and 15 arcsec. As mentioned in the Introduction, the bar of NGC~6032 presents some fine-structures. From Fig.~\ref{fig:light} it can be seen that the inner part of the bar resembles the barlens or X-shape morphologies described in \citet{2011MNRAS.418.1452L}, looking similar to a prominent bulge, and showing an exponential profile. 

(iii) Region 3: The outer bar. It is defined between 15 and 25 arcsec. Morphologically it can be distinguished as an elongated region located before the beginning of the spiral arms.

(iv) Region 4: The disc. Radially defined from 25 arcsec outwards. The region outside the outer bar.  

\begin{figure*}
	\includegraphics[width=\textwidth]{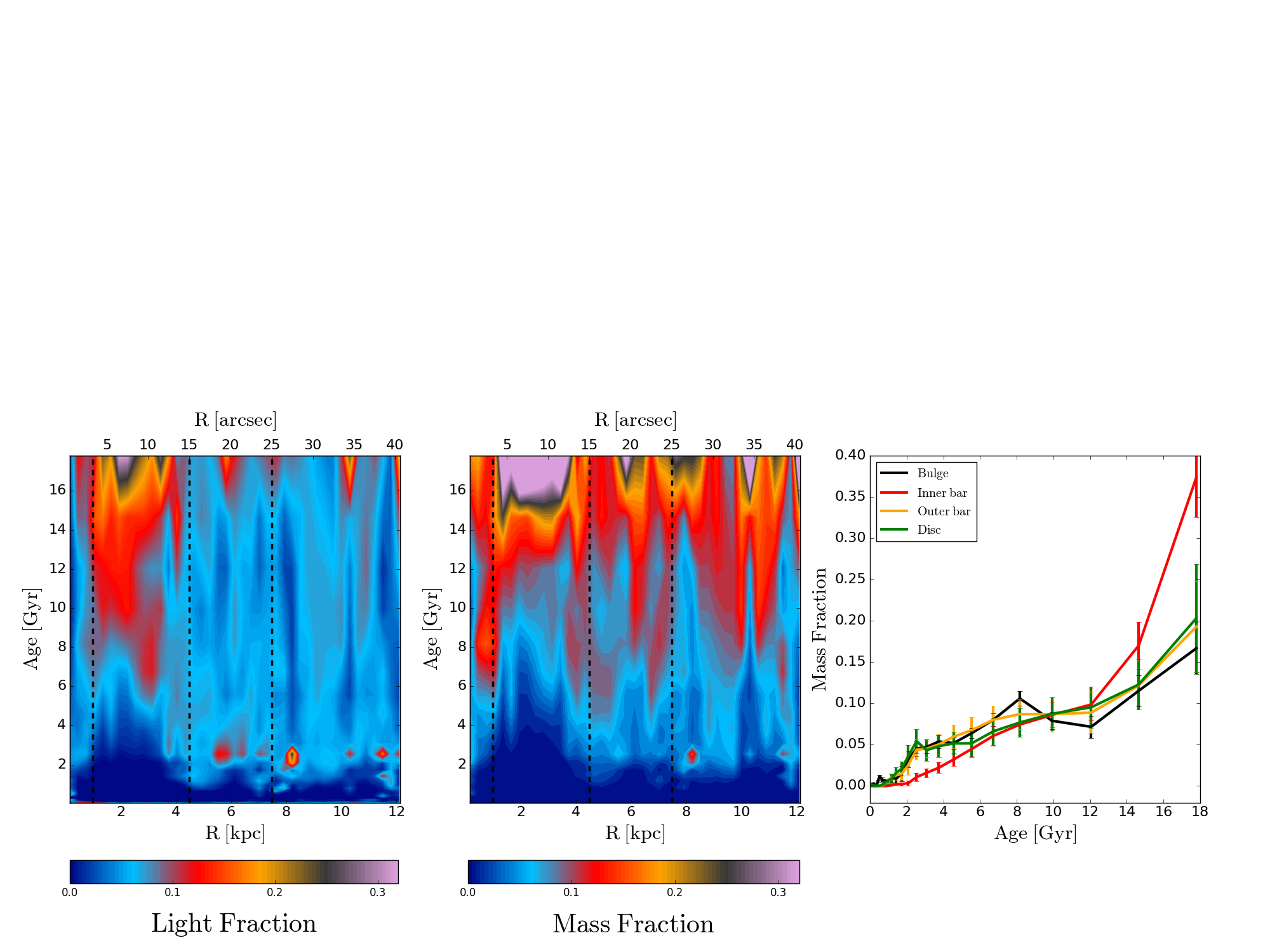}
    \caption{Left and middle panels: spatially resolved light- and mass-weighted Stellar Age Distribution (SAD), respectively. Blue colours imply no (or low) presence of stars of a given age (y axis) and at a given radius (x axis). The fraction of stars are normalised at every radius. The dashed vertical lines represent the analysed regions (see text). Right panel: average mass-weighted SAD for each of the analysed regions.}
    \label{fig:sfh}
\end{figure*}

From Fig.~\ref{fig:sfh} it can be seen that the stellar populations within the inner bar show a clear distinct SAD compared to that of the bulge, the outer bar, and disc regions. This fact is clearly shown when comparing the average SADs, where the outer bar and disc lines display similar SADs. The bulge also displays a similar SAD as these two regions with the exception of an excess of star formation at around 8~Gyrs ago. The figure also shows a common underlying old stellar population component for all the regions, describing the fact that very old stars are present all through the galaxy. In the normalised SAD plot this fact can be seen in the top part of the diagrams, where the fractions are relatively high for all regions. It is more clearly shown when looking at the averaged diagram on the right panel, where all the lines, one for each region, increase with age. The inner bar shows stars of ages down to 6 Gyrs old, while the bulge, outer bar, and disc regions present an excess of stars of younger ages. In other words, there is a deficit of stars younger than 4 Gyrs in the inner bar. With the exception of this inner bar region, as mentioned before, the rest of the galaxy shows the presence of stars of all ages. The outer bar and the disc show similar SADs.

\section{Comparison with {\bf a generic numerical simulation}}

\begin{figure}
	\includegraphics[width=\columnwidth]{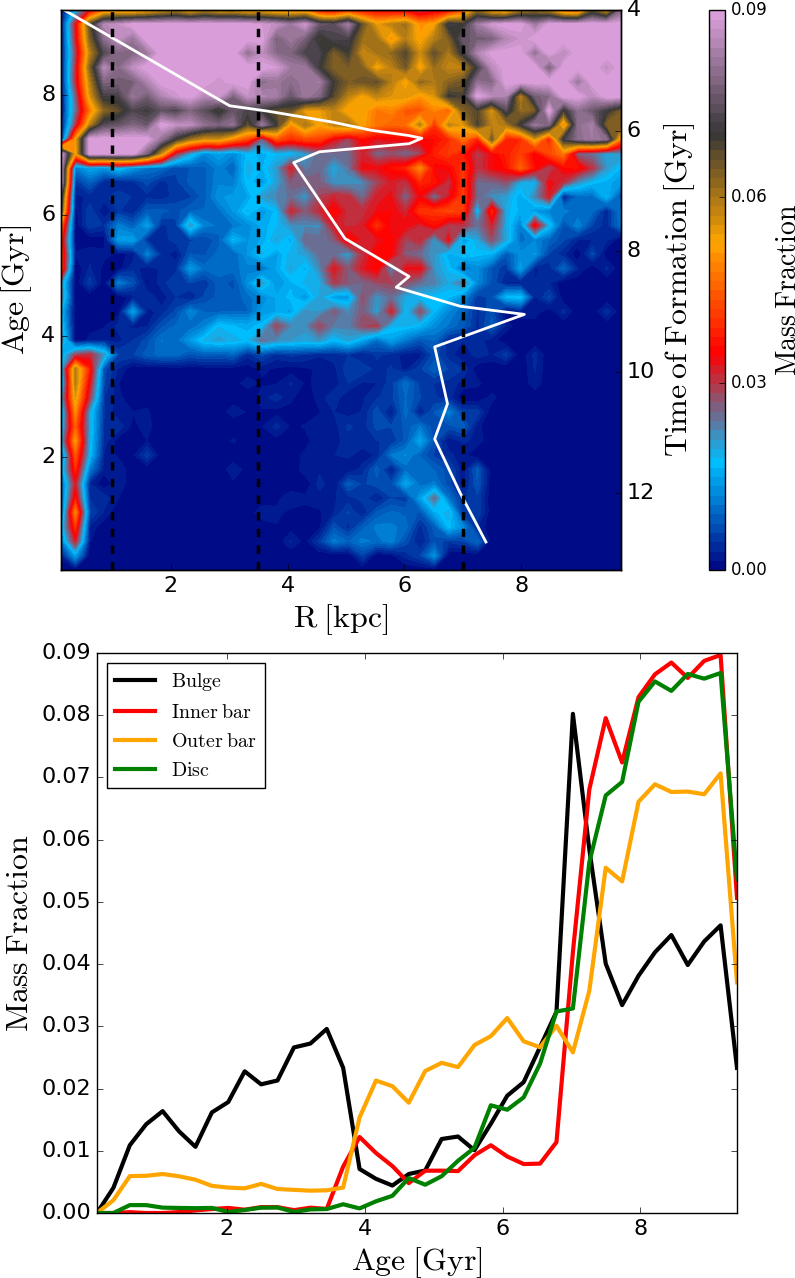}
    \caption{Top panel: radial distribution of stars (i.e. mass-weighted plot) created in the simulation at time 13.4 Gyr with age on the left and star formation time on the right. The bar size is outline with the solid white line. The buckling event occurs 7 Gyr ago creating most of the stars in the very inner region. Vertical lines delineate regions similarly to Fig.~\ref{fig:sfh}. Bottom panel: average mass-weighted SAD for the four regions, plotted in a similar way as Fig.~\ref{fig:sfh}}
    \label{fig:models}
\end{figure}

NGC~6032 is a clear example of a barred galaxy, and therefore, to interpret our findings from the CALIFA data on the stellar content we can compare them, as proof of concept, with a generic numerical simulation available within our group. The code used is that from the OWLS project \citep[][and references therein]{2010MNRAS.402.1536S} which includes $\it N$-body and SPH particles, star formation and stellar feedback, with thermal SN II feedback as described in \citet{2012MNRAS.426..140D}. The initial conditions are those from \citet{2011ApJ...734L..20M}, with 8\% of the disc material converted into SPH particles. The bar grows and becomes strong, then it buckles, weakens and grows again. In each of these phases the star formation is located at different regions of the bar. In Fig.~\ref{fig:models} we show in the top panel the fraction of stars at a given age (left axis) and a given radius, similarly to the observational data, and the time at which the stars form (right axis).The bottom panel shows the average mass-weighted SAD for the four analysed regions. The time and sizes should be interpreted in relative terms, as the simulation was not run nor scaled to be representative of NGC~6032. It can be seen from these two plots a main phase of star formation happening all through the disc before the bar is developed. These stars, together with those pre-existing the star formation period, would conform the oldest populations. Once the bar is formed it becomes strong, and then buckles and weakens again, from 6.5 to 7.3 Gyr ago. In this phase, driven by the dynamics, the star formation happens in the bar area and at the very centre. We expect an enhancement of the star formation in the central regions, i.e. the bulge, as gas is transported there associated to the strong bar and the buckling event. This phase seems to happen in this simulation around 7~Gyr ago. How much of this will happen in the centre depends on the position of the different resonances, which in turn depends on how the mass is distributed at the very centre. Later on, there is a long lasting phase (around 2~Gyr) of star formation outside the bar. Around 4 Gyr ago the bar has again a size of around 7 kpc and is forming stars in its outer as well as in the central parts. However, in what we call the inner bar there is some suppression of the star formation. Taking into account both, the evolutionary phases displayed by the simulated system and the similarities in the SFH from the generic simulation and the observed galaxy, NGC~6032, we can infer that the bar in this galaxy was formed more than 10 Gyr ago, as the morphological structure was already in place by then, an the excess of star formation observed at around 8~Gyr ago in the bulge region can be related to the buckling phase. 

We take further advantage of the simulation by exploring where the stars formed. The oldest stars suffer the strongest radial migration, a radial displacement on average of $\approx$~2 kpc. In particular, those created either at the outer part of the bar or outside the bar move on average $\approx$~1.5 kpc. In principle, new stars created at the centre stay in the centre, and those created in the bar outerparts stay around that region. 

\section{Discussion and Conclusions}

The relation between B/P bulges with the buckling instability of the bar has been observationally confirmed in recent studies (Erwin et al. 2016). However, establishing the moment at which this event occurs and the subsequent time evolution is crucial to understand the present and past observed frequency and the properties of B/P galaxy bulges, including that of our Milky Way. In this letter we characterise the stellar content of the different morphological components of NGC~6032, focusing on the bar and central structures of NGC~6032. The bulge of NGC~6032 resembles the X-shape and barlens bulges found in galaxies at low or intermediate inclinations that are possibly associated to the presence of buckled bars \citep[e.g.][]{2014MNRAS.444L..80L, 2015MNRAS.454.3843A}. The stellar age distributions observed in NGC~6032 are closely linked to the present morphological structures, suggesting that these regions have been in place for a long time without major stellar mass redistribution. Following this fact, we can interpret the observed SADs as the result of the star formation histories that occurred in NGC~6032. This interpretation of the observations allows us to directly compare them with numerical simulations. The main discussion points and conclusions regarding the comparison between the SFHs of NGC~6032 and the results from the numerical simulation are summarised below:
\begin{itemize}

\item Constraints on the bar formation and buckling period: The enhancement of star formation in the inner bar, as compared to the neighbour regions, until around 6 Gyrs ago suggests, based on the new simulation described above, a relation with the bar formation and main growth, right before the buckling event and the quenching of star formation later on. In this way we can estimate from Fig.~\ref{fig:sfh} the buckling occurring around 8 Gyrs ago. We can also infer the formation of the stable disc and the bar at least before 10~Gyrs ago following a quiet evolution since then, i.e., without major mergers disrupting the disc. In this picture, the underlying disc forms and as it becomes unstable forms the bar that later buckles to form the B/P bulge at around 8 Gyrs. 

\item After the event of the B/P formation, the bar grows again mainly from disc material, given that the outer bar presents a star formation history similar to that of the disc. For the first time, we present observational evidence of the bar growing from disc material. 


\item Subsequent star formation in the bar region: It is interesting to notice that the star formation in the inner regions of the galaxy goes on despite the quenching in the star formation of the inner bar, suggesting that the bar can act as a conveyor belt, transferring  material to the centre from the disc without forming stars within it. From the theoretical point of view, it could be clearly explained by the fast gas inflow triggered by the bar towards the inner parts through the $\it{dust lanes}$. The accumulation of cold material in the centre triggered therefore the star formation. Meanwhile outside this region, in the disc, there are regions of recent star formation. This is in agreement with other recent observational studies such as \citet{Consolandi16} where he concludes that bars are redder structures with respect of their discs.

\item As already mentioned, the fact that different star formation histories are so closely linked to the present observed morphological fine-structures suggests that the bar in NGC~6032 has been in place for a long time. It has been previously argued \citep{2002A&A...392...83B} that bars of late-type galaxies, i.e. presenting gas in their discs, would weaken and last for only a few Gyrs before reappearing again. Although some observational evidence of bars being long-lasting for early-type galaxies has been provided \citep[e.g.][]{2009A&A...495..775P, 2011MNRAS.415..709S, 2015A&A...584A..90G, 2015MNRAS.446.2837S}, this is the first time that it is clearly shown for a prototypical late-type, Sb, galaxy.

\end{itemize}

We show, for the first time, that the analysis of the star formation histories of the different structures of NGC~6032 provide sufficient evidence to trace the moment of bar formation and its growth. We also constrain the buckling phase of the bar and their subsequent evolution for the nearby galaxy NGC~6032. We trace the bar formation at around 10 Gyrs and the buckling phase of NGC~6032 possibly happening at around 8~Gyrs ago. We conclude that, from that moment, the bar grows from disc material and it does not significantly form stars while transporting material to the central parts. These results are supported by recent numerical simulations. The same analysis on a bigger sample of galaxies will help us to generalise these results to other galaxies and to shed some light into the formation of stellar discs.

\section*{Acknowledgements}
We thank the referee for the useful discussion and comments that have greatly helped to improve this Letter.This  Letter  is  based  on  data  from  the  CALIFA  Survey,  funded  by  the  Spanish  Ministry of Science (grant ICTS-2009-10) and the Centro Astron\'omico Hispano-Alem\'an. This work has been supported by the Spanish Ministry of Science and Innovation under grants AYA2016-77237-C3-1-P, and Consolider-Ingenio CSD2010-00064; and by the Junta de  Andaluc\'ia  (FQM-108), and the AYA2014-53506-P grant, funded by the 'Ministerio de Econom\'ia y Competitividad' and by the 'Fondo europeo de desarrollo regional FEDER'. RAM acknowledges support by the Swiss National Science Foundation. We acknowledge the contribution of Teide High-Performance Computing facilities. TeideHPC facilities are provided by the Instituto Tecnol\'ogico y de Energ\'ias Renovables (ITER, SA). URL:http://teidehpc.iter.es. IMV has been partially supported by MINECO AYA2014-583308-P. SFS thanks the Conacyt programs 180125 and DGAPA IA100815.




\bibliographystyle{mnras}
\bibliography{bibliography} 








\bsp	
\label{lastpage}
\end{document}